\documentclass{jltp}

\usepackage{graphicx} % uncomment this line to include the graphicx package

\newcommand\ket[1]{|#1\rangle}
\title{Ferromagnetism in a Realistic Two-Band Model:\thanks{Dedicated
to Peter W\"olfle on the occasion of his 60th birthday.}\\ A Slave-Boson Study}

\author{Raymond Fr\'esard$^{\S\dag}$ and Mathieu Lamboley$^{\S\ddag}$}

\address{\S Institut de Physique, Universit\'e de Neuch\^atel,
A.-L. Breguet 1, 2000 Neuch\^atel,\\
Switzerland\\
\dag Laboratoire Crismat, UMR 6508 du Centre National de la
Recherche \\
Scientifique et de l'Institut des Sciences de la Mati\`ere et
du Rayonnement, \\
6, Bld. du Mar\'echal Juin, 14050 Caen Cedex, France\\
\ddag Laboratoire de G\'enie M\'edical, EPFL, 
PSE A, 1015 Lausanne, Switzerland}

\runninghead{Raymond Fr\'esard and Mathieu Lamboley}{Ferromagnetism in
a Realistic Two-Band Model.} 

\begin{document}

\maketitle

\begin{abstract}
Using a slave boson representation of multi-band Hubbard models, we
investigate a two-band model relevant to layered perovskites in the
vicinity of half-filling. Beside the strong influence of the Hund's
rule coupling, we obtain that the phase diagram separates into two
regions: a weak to moderate coupling region where the effective mass
is weakly renormalized, and a strong coupling regime where it is
strongly renormalized. The transition between these two regimes is
very sharp. It takes place in a (vanishingly) small domain. A
ferromagnetic instability in only found in the strongly correlated
regime, and is triggered by the Hund's rule coupling. The results are
compared to La-doped layered ruthenates.

\end{abstract}

\section{INTRODUCTION}
The richness of the phase diagram of the Ruddlesden-Popper series of
the ruthenates makes it
a fascinating topics of current solid state physics. Indeed
single-layered ruthenates can be a superconductor
(Sr$_{2}$RuO$_{4}$) as discovered by Maeno {\it et al.}\cite{Maeno94}
or an anti-ferromagnetic insulator 
(Ca$_{2}$RuO$_{4}$), 
while multi-layered compounds may be ferromagnetic metals
(SrRuO$_{3}$ and possibly Sr$_{3}$Ru$_{2}$O$_{7}$) or an anti-ferromagnetic
metal (CaRuO$_{3}$). This
richness is particularly amazing when one observes that all
these compounds are, from a theoretical point of view,
remarkably similar. Indeed Ca and Sr are divalent cations, the 
electron-electron interaction is 
strongest on the Ru atoms, which are common to all compounds,
and the lattice structure does not drastically change when
scanning through this family. Nevertheless these small
changes must be responsible for the variety of the ground
states. This thus calls for an accurate description of the
electronic band structure, for instance on the level of the tight binding
approximation. Note that Ca$_2$RuO$_4$ appears to be a Mott insulator below
300 K. Indeed, transport\cite{Cao97} and optical\cite{Puchkov98}
measurements reveal a narrow gap of about 0.2 eV. It orders
magnetically below $T_N = $100 K, and therefore the insulating state
may not be attributed to the magnetic ordering. 

Another striking feature is provided by La doping (or other trivalent
cations) of both Sr- and Ca-based single layered compounds: in both
cases a few percent doping turns these systems into
ferromagnets\cite{Cao00,Cao00b}. 
For the Ca-based compound, the Curie temperature is
about 100 K, while 
substitutions of Ca with other trivalent dopants, such as Y, Ce or Pr
result in ferromagnetic states with Curie temperatures varying from 120
to 200 K\cite{Cao00,Cao00c}. For the Sr-based compound, La substitution
results into a ferromagnetic state, 
with a Curie temperature of 40 K at a concentration of 4 percent of
dopant\cite{Cao00b}. 

Instead of investigating the nature and  the cause of the
superconductivity  in 
Sr$_2$RuO$_4$, which is still a controversial
matter\cite{Eschrig01,Maeno00,Sigrist99,Baskaran96,Rice95}, we rather  
concentrate on the Fermi liquid parameters and the magnetic structure
of the entire series. The paper is organized as follows: we first
introduce a model meant to capture the main feature of this series on
a minimal model level. It includes the details of the band structure,
inspired by a proposition from Noce and Cuoco\cite{Noce99}  which is
compatible with the experimentally determined Fermi
surface\cite{Mackenzie96} and spectroscopic data\cite{Schmidt96}. With
minimal 
changes it can be applied to other perovskites such as vanadates and
chromites. 
Secondly we explain how we are dealing with the strong Coulomb and
Hund's rule couplings. Thirdly we present numerical results pointing
towards a picture where both Ca-, and Sr-based ruthenates derive from a
Mott insulator. The former system appears to be a Mott insulator at
commensurate filling, while the second one should rather be seen as a
self-doped Mott insulator. This is due to a partial filling of all
sub-bands. 

\section{THE MODEL}
\subsection{Band Structure}
The band structure of Sr$_{2}$RuO$_{4}$ has been determined in
the LDA by Oguchi\cite{Oguchi95} and Singh\cite{Singh95}. Later on
it turned out that it can 
accurately be reproduced by a simple tight binding model by 
Noce and Cuoco\cite{Noce99}. The bands which are crossing the Fermi
energy are  
involving the three 4d t$_{2g}$ orbitals of Ru, which are
hybridizing with the 2p orbitals of O. They are filled with 16
electrons per RuO$_{2}$ unit. Since the RuO$_{6}$ octahedra 
are elongated along the z-direction, the d$_{xy}$ orbital 
energy level is lying lower than the other two. From the LDA 
calculation, the corresponding band is nearly filled.  
In the following it will be neglected, except for possibly 
contributing to self-doping effects. The d$_{x\,(y)z}$ orbitals 
hybridize only with the p$_{x\,(y)}$ orbital of the %x(y)
neighboring O along the z-axis, and with the p$_z$ orbital along
the x (y)-axis. On top, the latter two orbitals weakly couple to one
another as well. It numerically turns out that the dispersion in the 
z-direction is small, and will therefore be neglected. The resulting
dispersions are given by the eigenvalues of: 
\begin{equation}\label{ie}
H_{z} = \sum_{k,\sigma}\Psi^{\dagger}_{k,\sigma}
\left(
%\begin{array}{cccccc}
\begin{array}{cccc}
\epsilon_{d} & 0 & -ie_{k} & 0 \\
0 & \epsilon_{d} & 0 & -if_{k}  \\
ie_{k} & 0 & \epsilon_{p} & a_{k}  \\
0 & if_{k} & a_{k} & \epsilon_{p}  
\end{array}
\right)
\Psi_{k,\sigma}\, ,
\end{equation}
with $\Psi^{\dagger}_{k,\sigma} = \left( d_{xz}^{\dagger},
d_{yz}^{\dagger},p_{1z}^{\dagger},p_{2z}^{\dagger}\right)_{k,\sigma}$, 
and $a_{k} =
-4t_{3}\cos{\left(\frac{k_{x}a}{2}\right)}\cos{\left(\frac{k_{y}a}{2}\right)}$,
$e_{k} = 2t_{5}\sin{\left(\frac{k_{x}a}{2}\right)}$, and $f_{k} =
2t_{5}\sin{\left(\frac{k_{y}a}{2}\right)}$. 
%
%\begin{equation}\label{if}
%
%\Psi^{\dagger}_{k,\sigma} = \left( d_{xz}^{\dagger},
%d_{yz}^{\dagger},p_{1z}^{\dagger},p_{2z}^{\dagger}\right)_{k,\sigma} \, ,
%
%\end{equation}
%
%\begin{eqnarray}\label{ik}
%
%a_{k} &=&
%-4t_{3}\cos{\left(\frac{k_{x}a}{2}\right)}\cos{\left(\frac{k_{y}a}{2}\right)}\,
%, \nonumber\\ 
%e_{k} &=& 2t_{5}\sin{\left(\frac{k_{x}a}{2}\right)}\, , \\
%f_{k} &=& 2t_{5}\sin{\left(\frac{k_{y}a}{2}\right)}\, . \nonumber
%
%\end{eqnarray}
%
Here $ -\frac{2\pi}{a}\leq k_{x}\, (k_{y}) \leq \frac{2\pi}{a}$ and
$a$ represents the lattice constant (the Ru-Ru distance). It is 
from now on set to unity. The LDA dispersions is best reproduced with
the parameter set $t_{3} = 0.1\, eV$, $t_{5} = 0.85\, eV$,
$\epsilon_{p} = -2.4\, eV $ and $\epsilon_{d} = -0.9\, eV$. Integrating
out the oxygen bands, and neglecting their frequency dependence, yields
the effective model:
\begin{eqnarray}\label{ip}
H_{0} &=&  \sum_{k,\sigma}
\left(
d_{xz,\sigma}^{\dagger},d_{yz,\sigma}^{\dagger}
\right)
\left(
\begin{array}{cc}
\tilde{e}_{k} & \tilde{a}_{k}
\\ [0.3cm]
\tilde{a}_{k} & \tilde{f}_{k} \\
\end{array}
\right)
\left(
\begin{array}{c}
d_{xz,\sigma}\\d_{yz,\sigma}
\end{array}
\right)\, ,
\end{eqnarray}
up to a constant energy shift. Here we introduced: $\tilde{a}_{k} =
-4t'\sin{k_{x}}\sin{k_{y}}$, $\tilde{e}_{k} = t\cos{k_{x}}$, and $\tilde{f}_{k} = t\cos{k_y}$.
The effective hoppings $t$ and
$t'$ are related to the original parameters by $t = 2
\frac{t_5^2}{\epsilon_{p}}$ and $t'=t_3 \left(\frac{t_5}{\epsilon_{p}}
\right)^2$. Note that the two-dimensional character of the dispersions
\begin{equation}\label{disp}
E_{k,\nu,\sigma} = %a_s + 
\frac{1}{2}\left( \tilde{e}_{k}  +
\tilde{f}_{k} \right) + \frac{1}{2}\nu\sqrt{\left(\tilde{e}_{k}  -
\tilde{f}_{k}\right)^{2} + 4\tilde{a}_{k}^2} \, ,
\end{equation}
with $\nu=\pm1$, purely follows from the off-diagonal terms. These two
eigenvalues 
yield the dispersion one 
obtains in LDA calculation\cite{Noce99}. The resulting density of
states is symmetrical with respect to half band filling, where it has
a minimum. The band width $W$ is 1.2~eV.
\subsection{Interaction Terms}
We have modeled the Coulomb interaction using a Hubbard $U$-term and 
Hund's rule coupling term $J_H$-terms, yielding  the local
Hamiltonian:  
\begin{eqnarray}\label{ham}
H_{int} &=& U \sum_{i,\sigma,\rho'<\rho}n_{i,\rho,\sigma}
n_{i,\rho',\sigma} 
\nonumber\\[0.3cm]
&+&U_1\sum_{i,\rho'\neq\rho}n_{i,\rho,\uparrow}n_{i,\rho',\downarrow}
+U_3 \sum_{i,\rho}
n_{i,\rho,\uparrow}n_{i,\rho,\downarrow}\, ,
\end{eqnarray}
where $\sigma$ ($\rho$) is a spin (band) index. The
relation between the coefficients 
\begin{equation}\label{hund}
U_n \equiv U+nJ_H
\end{equation}
holds for perfect cubic symmetry\cite{Sugano70l,Bunemann98}, otherwise
one should 
work with the matrix elements $F_0$, $F_2$, and $F_4$\cite{Fresard97}. In
the following we nevertheless stick to the minimal model using the
relation Eq. (\ref{hund}). 
This amounts to deal with the parameters
$U$, $J_H$, $t$, and $t'$ (with $t$ setting the energy
scale). Since the ruthenates (Sr$_{1-x}$Ca$_x$)$_2$RuO$_4$ may be Mott
insulators (depending on the value of $x$), perturbative calculations
may be problematic. Here we resort to a slave boson
approach\cite{Dorin93,Fresard97,Hasegawa97}, which is the
generalization of the  
Kotliar and Ruckenstein representation of the Hubbard
model\cite{Kotliar86}, to the multi-band model. 
At this stage one may wish 
to introduce a rotation invariant formulation in the spirit of Li {\it
et al.} or 
Fr\'esard and W\"olfle's representations\cite{Li89,Fresard92}. In the
one-band model, such a representation  turned out to be necessary to
obtain the correct contribution of the spin fluctuations to the
specific heat on the one-loop level\cite{Li89}. It also appeared very
useful when investigate incommensurate
states\cite{Fresard91b,Fresard91,Arrigoni91}. In the systems under
consideration, perfect cubic symmetry is absent, and therefore
rotational symmetry in spin space is not a requirement. Thus the local
model Eq. (\ref{ham}) can be viewed as a model in its own right,
despite of lacking rotational invariance.
Note that the tightly related Gutzwiller Approximation has been
suitably generalized to deal with rotationally symmetrical
models\cite{Bunemann98}. 
Manifestly anti-ferromagnetic
solutions are playing a role as well, as discussed by
Hasegawa\cite{Hasegawa97b}. This is however a less puzzling issue than
the occurrence of ferromagnetism, and is therefore not investigated
here. Since there is no van Hove singularity at half-filling, we
expect that an anti-ferromagnetic instability can only occur above a
finite (and substantial) $U$, as it happens in the $t-t'-U$
model\cite{Yang00} or in the one band model on the hexagonal
lattice\cite{Fresard95}. 
Recently Hasegawa\cite{Hasegawa00} showed that orbitally ordered
states in this model seem to rather appear for negative $J_H$.

In the slave boson approach to the multi-band model one
introduces one slave boson field for each of the sixteen atomic
configurations, and four auxiliary fermionic fields $f_{\alpha}$
($\alpha$ being a composite spin and band index) as :
\begin{eqnarray}\label{bosonsmoi}
%
%\ket{0}&=& e^{\dagger} \ket{vac}\, , \nonumber\\[0.3cm]
%
\ket{\alpha}&=& p_{\alpha}^{\dagger} f_{\alpha}^{\dagger} \ket{vac}\, , \nonumber\\[0.3cm]
\ket{\alpha\, , \alpha^{'}}&=& d_{\alpha \alpha^{'}}^{\dagger}
f_{\alpha}^{\dagger} f_{\alpha^{'}}^{\dagger} \ket{vac}\, , \\[0.3cm]
\ket{\alpha\, , \alpha^{'}\, , \alpha^{''}} &=&
t_{\alpha \alpha^{'} \alpha^{''}}^{\dagger}
f_{\alpha}^{\dagger} f_{\alpha^{'}}^{\dagger}
f_{\alpha^{''}}^{\dagger} \ket{vac}\, . \nonumber
%\\[0.3cm]
%
%\ket{\alpha\, , \alpha^{'}\, , \alpha^{''}\, , \alpha^{'''}}&=&
%q^{\dagger}f_{\alpha}^{\dagger}  f_{\alpha^{'}}^{\dagger}
%f_{\alpha^{''}}^{\dagger} f_{\alpha^{'''}}^{\dagger}\ket{vac}\,
%.\nonumber
%
\end{eqnarray}
On top of those, there is a boson $e$ related to empty sites, and a
boson $q$ related to four-fold occupied sites. The original fermionic
operators $c^{\dagger}_{i \alpha}$ are 
expressed in terms of the auxiliary fields as $c^{\dagger}_{i \alpha}
= z^{\dagger}_{i \alpha} f^{\dagger}_{i \alpha}$ 
%
%
%\begin{equation} \label{eqc}
%c^{\dagger}_{i \alpha} = z^{\dagger}_{i \alpha} f^{\dagger}_{i \alpha}
%\end{equation}
with $z^{\dagger}$ given by Eq. (4-5) of Ref. \onlinecite{Fresard97}, while the
constraints linking the auxiliary fields are given by Eq. (6) of
Ref. \onlinecite{Fresard97}. 
\begin{figure}[t]
\vspace*{7mm}
\centerline{\includegraphics[height=2.4in]{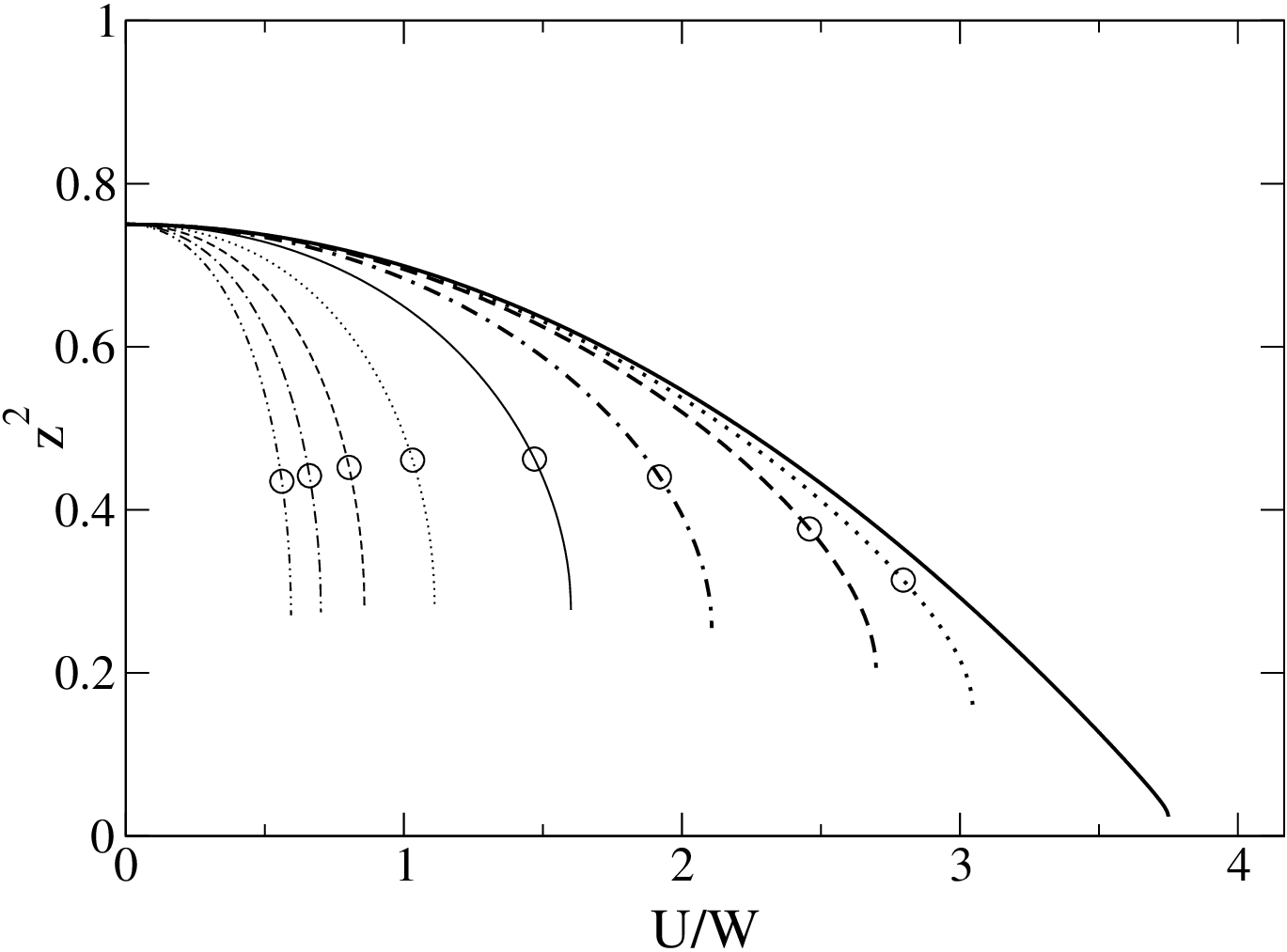}}
\caption{Effective mass renormalization at $\rho=2$ for $J_H/U$= 0 (thick
full line), 0.01. (thick dotted line), 0.02 (thick dashed line),
0.05 (thick dashed-dotted line), 0.1 (thin
full line), 0.2. (thin dotted line), 0.3 (thin dashed line),
0.4 (thin dashed-dotted line) and 0.5 (thin dashed-dotted-dotted
line). The circles indicate the location of the first order transition.}  
\label{fig:z_half}
\end{figure}
Altogether, after having integrated out the
fermions, we obtain the grand potential in saddle-point approximation as :
\begin{eqnarray}\label{grandpot}
\Omega  &=&
U\sum_i\left(\sum_{\alpha<\alpha^{'}}d_{i,\alpha \alpha'}^{2} + 3
\sum_{\alpha < \alpha^{'} <
\alpha^{''}}t_{i,\alpha\alpha^{'}\alpha^{''}}^{2} 
+ 6 q_i^{2}\right) \nonumber\\[0.3cm]
&+& J\sum_i \left(\sum_{\sigma} d_{i,xz \sigma, yz -\sigma}^{2} + 
3\sum_{\rho} d_{i,\rho \uparrow, \rho \downarrow}^{2} +
4\sum_{\alpha < \alpha^{'} < \alpha^{''}}
t_{i,\alpha\alpha^{'}\alpha^{''}}^{2} +  
8q_{i}^{2}\right)\nonumber\\[0.3cm]
&+& \sum_i\lambda_i\left(
e_{i}^{2}+\sum_{\alpha}p_{i\alpha}^{2}+\sum_{\alpha<\alpha^{'}}d_{i,\alpha\alpha^{'}}^{2}+
\sum_{\alpha < \alpha^{'} < \alpha^{''}} t_{i,\alpha\alpha^{'}\alpha^{''}}^{2}+q_{i}^{2}-1\right)\\[0.3cm]
&-& \sum_{i,\alpha}\beta_{i,\alpha}\left(p_{i,\alpha}^{2}+
\sum_{\alpha^{'}}d_{i, \alpha \alpha^{'}}^{2}+
\sum_{\alpha^{'} \alpha^{''}}t_{i,\alpha\alpha^{'}\alpha^{''}}^{2} +
q_{i}^{2}\right)\nonumber\\[0.3cm]
&- & \frac{1}{\beta}\sum_{k,\nu,\sigma}\ln\left({1+e^{-\beta
E_{k,\nu,\sigma}}}\right)\nonumber
\end{eqnarray}
In a paramagnetic or a ferromagnetic phase, the dispersions are given
by :
\begin{eqnarray}\label{disprenorm}
E_{k,\nu,\sigma} &=& \frac{1}{2}\left[ 
%2a_s 
\beta_{xz\sigma}+\beta_{yz\sigma} -2\mu 
+ \left(  z_{xz \sigma}^{2}\tilde{e}_{k}  +
z_{yz \sigma}^{2} \tilde{f}_{k} \right) + \right.\nonumber \\
& & + \left. \nu\sqrt{\left(z_{xz \sigma}^{2}\tilde{e}_{k}  - z_{yz \sigma}^{2}
\tilde{f}_{k}\right)^{2} +
4 z_{xz \sigma}^{2}z_{yz \sigma}^{2} \tilde{a}_{k}^{2}} \, \right]
\end{eqnarray}
where the dependence on $\sigma$ is only effective in the
ferromagnetic phase. 
\begin{figure}[t]
\vspace*{7mm}
\centerline{\includegraphics[height=2.5in]{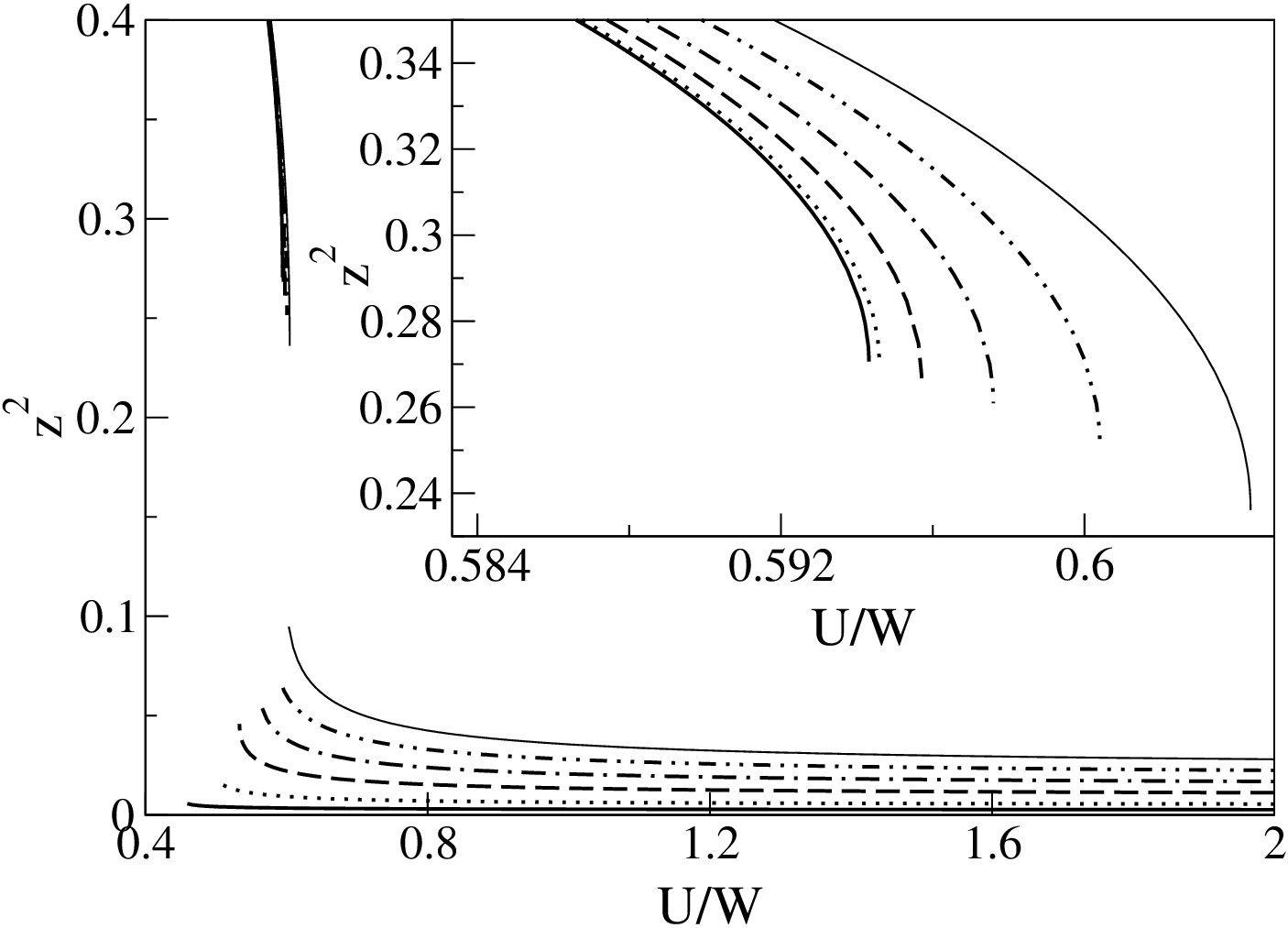}}
\caption{Effective mass renormalization off half-filling for $J_H/U$=
0.5  and $\rho=2.005$ (full line), 2.01. (dotted line), 2.02 (dashed line),
2.03 (dashed-dotted line), 2.04 (dashed-dotted-dotted
line), and 2.05 (thin full line). Inset: Blow up of the ``metallic''
solutions with the same parameters.}  
\label{fig:z_dop_2}
\end{figure}
Though the resulting mean-field equations only
correspond to the 
saddle-point of an action, the latter gets variationally controlled in
the limit of large dimensions\cite{Bunemann97b}. It would be highly
desirable to reach the degree of accuracy and reliability of the
conserving approximations developed by the Karlsruhe group for
impurity models\cite{Kroha01,Kroha92,Kroha97,Kroha98}, but to date it
appears difficult to extend them to lattice models. 
\begin{figure}[t]
\vspace*{7mm}
\centerline{\includegraphics[height=2.5in]{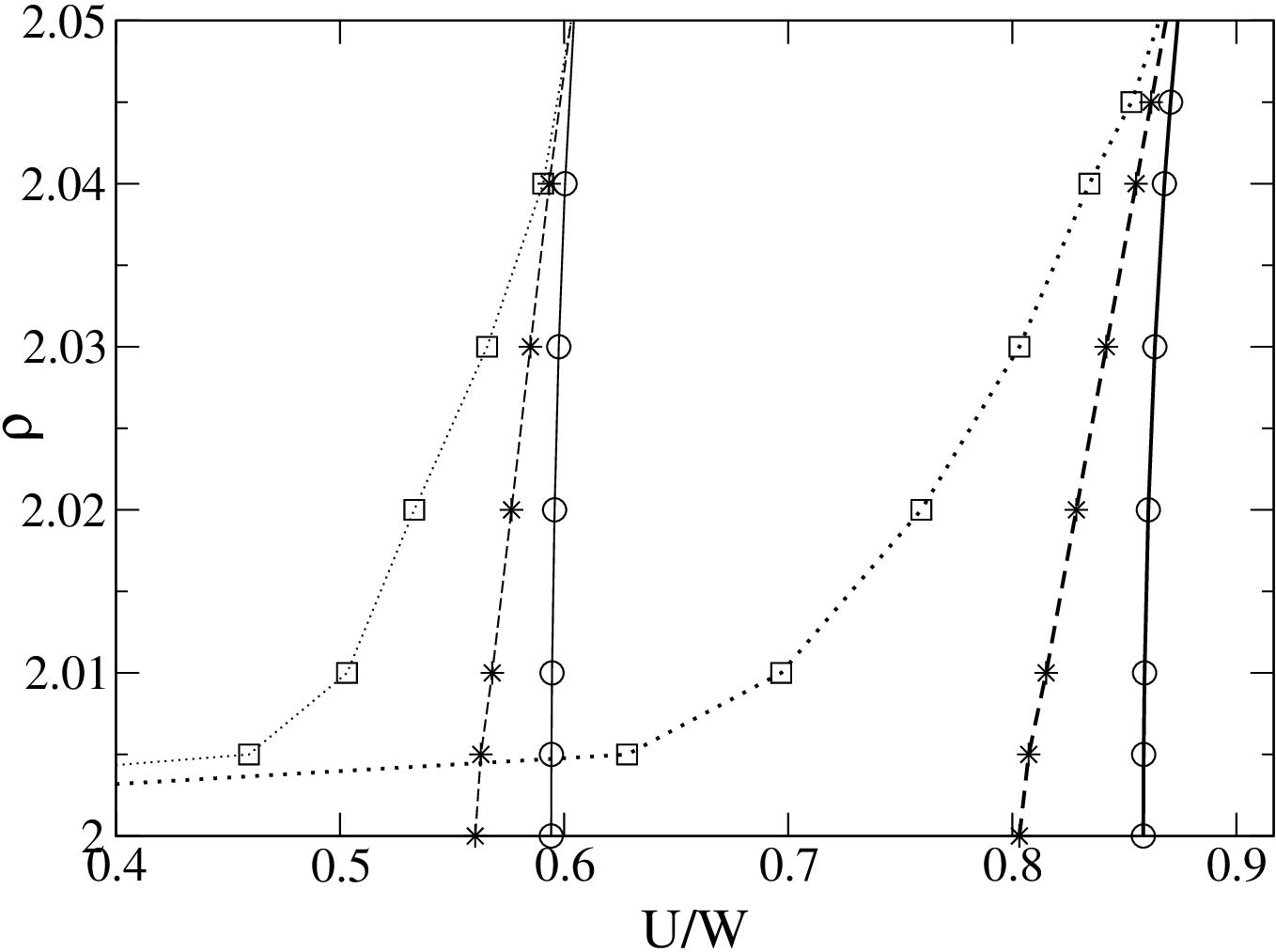}}
%
%\framebox[5in]{\rule[1.125in]{0in}{1.125in}}
%\makebox[5in]{\rule[1.125in]{0in}{1.125in}}
\caption{Coexistence regions of the paramagnetic phases for $J_H/U$= 0.5
(thin lines) and 0.3, (thick lines). The dashed lines locate the first
order transition lines, the dotted lines $U_{c1}$, and the full
lines $U_{c2}$. All lines are guide for the eyes, only.}  
\label{fig:coex}
\end{figure}
Nevertheless mean-field
calculations turned out to compare well with numerical
simulations\cite{Fresard91b}, even when calculating the charge
structure factor\cite{Zimmermann97}. 
\section{NUMERICAL RESULTS}
The saddle-point equations have been solved on a 800 $\times$ 800
lattice, at a temperature T = t/1000.  
Since the electronic density in the ruthenates under study is $\rho
\sim 2$,
we neglected four-fold occupancies and empty configurations. This
approximation is justified in the vicinity of the Mott
transition. However it breaks down for densities above three (below
one), and for weak coupling, where our
results should be taken with care. In particular it results into
having a finite effective mass renormalization in the non-interacting
limit, which should not be the case.
As obtained by
B\"unemann et al.\cite{Bunemann97,Bunemann98}, and Klejnberg and
Spalek\cite{Klejnberg98},  the Hund's rule coupling
has a strong influence on the Mott transition. While the latter is
second order for $J_H = 0$ and $\rho=2$, or for any $J_H$ for $\rho =
1$ or 3, it becomes first order for finite $J_H$ at half-filling as
shown on Fig. \ref{fig:z_half}. The metallic solution of the 
saddle-point equations ceases to exist at a critical value $U_{c2}$, 
without corresponding to a diverging effective mass. Moreover the 
effective mass is at most renormalized by a factor five for $J_H/U \geq 
0.01$, in contrast to the one-band case. On top of this metallic
solution, there is an insulating paramagnetic
solution characterized by a vanishing value of all bosons, but $d_{xz
\uparrow, yz \uparrow}$ and $d_{xz \downarrow, yz \downarrow}$, and
therefore a diverging effective mass (for finite $J_H$). 
\begin{figure}[t]
\vspace*{7mm}
\centerline{\includegraphics[height=2.5in]{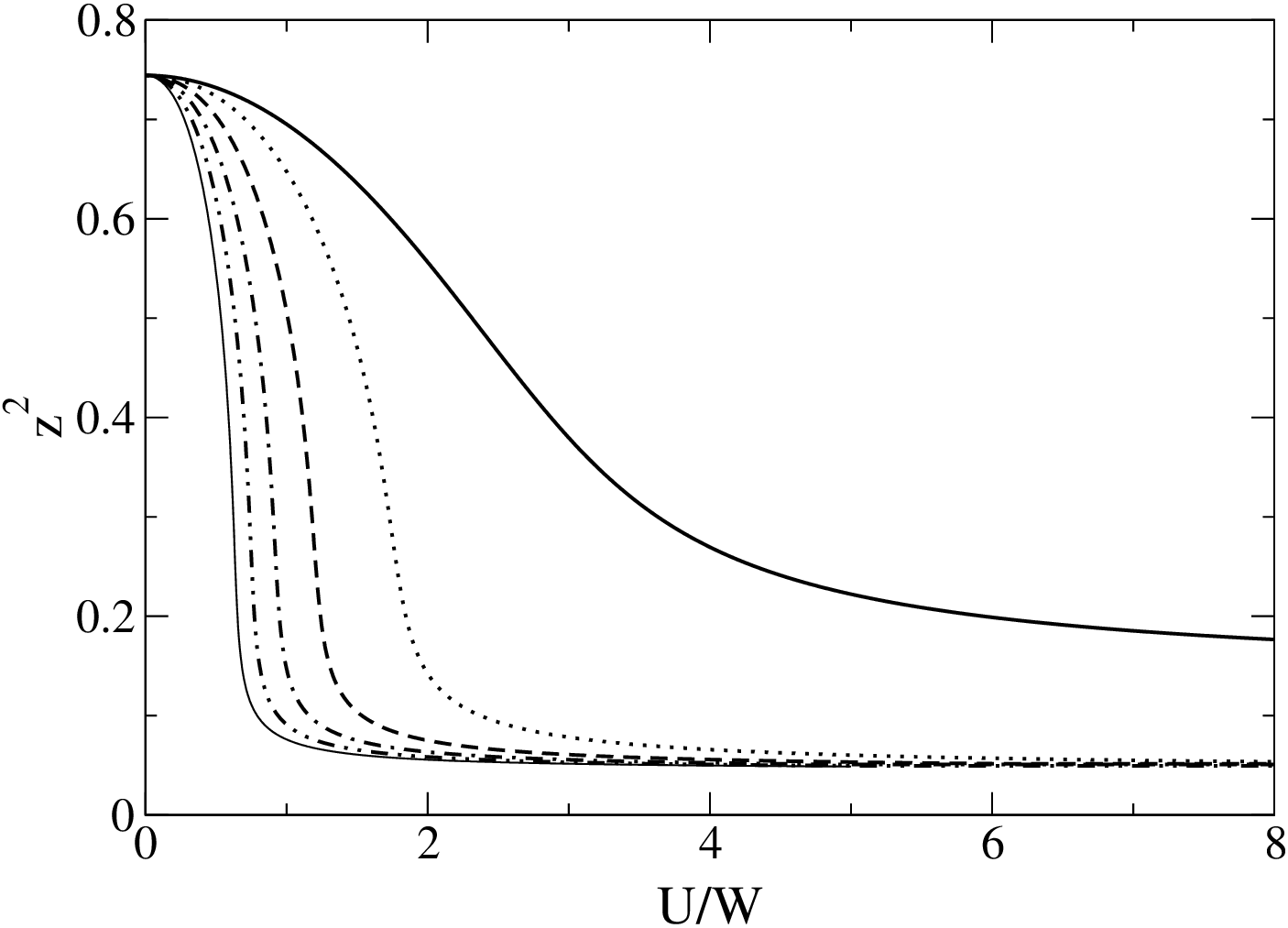}}
\caption{Effective mass renormalization at $\rho=2.1$ for $J_H/U$=
0 (full line), 0.1 (dotted line), 0.2 (dashed line),
0.3 (dashed-dotted line), 0.4 (dashed-dotted-dotted
line), and 0.5 (thin full line). }  
\label{fig:z_dop0p1}
\end{figure}
It extends down to $U_{c1} = 0$.
We also note that $U_{c2}$ is
slightly larger than $U_c$, where the energy of the metallic and 
insulating solutions coincide. As a result the effective mass 
renormalization is even more modest in the metallic phase.  

Once the system is doped the situation changes little by little. For small
electron doping, the first order transition remains but gradually 
vanishes with increasing electron concentration as shown on 
Fig. \ref{fig:z_dop_2}. The metallic solution is only modestly
affected, except for that it allows for decreasing values of $z$ in the
vicinity 
of the Mott transition. The insulating solution becomes metallic under
electron doping, and  the truly insulating state is only found for
integer fillings.  
However the effective mass renormalization remains very large,  and
accordingly the quasi-particle residue is small. Under these 
circumstances, the system may well be strongly influenced by other 
interaction terms, as reviewed by Vollhardt et al.\cite{Vollhardt97}, 
or disorder effects. This may  
explain why many transition metal oxides remain insulating even upon 
substantial doping, such as La$_{1-x}$Ca$_x$VO$_3$\cite{Nguyen95}.
  
When coming from large U, the "insulating" solution is found to exist 
down to a critical value $U_{c1}$, which is smaller than 
$U_{c2}$. Beside the fact that $U_{c1} = 0$ at half-filling  for any
$J_H$, it is a complex function of density and $J_H$ otherwise. It is
shown on Fig. \ref{fig:coex} for $J_H/U = 0.3$ and 0.5; together with
$U_c$ and $U_{c2}$. The coexistence region is seen to be largest for 
$J_H/U = 0.3$. It gets somewhat smaller for larger ratio of $J_H/U$, but 
its size drops abruptly for $J_H/U \rightarrow 0$. 
For larger doping the effective mass renormalization is getting a 
smooth function of $U$, as displayed on Fig. \ref{fig:z_dop0p1}. Again 
it strongly depends on $J_H$, either in the vicinity  
of the Mott transition since its location depends on $J_H$, or when it
is small, even for large $U$. 
\begin{figure}[t]
\vspace*{7mm}
\centerline{\includegraphics[height=2.5in]{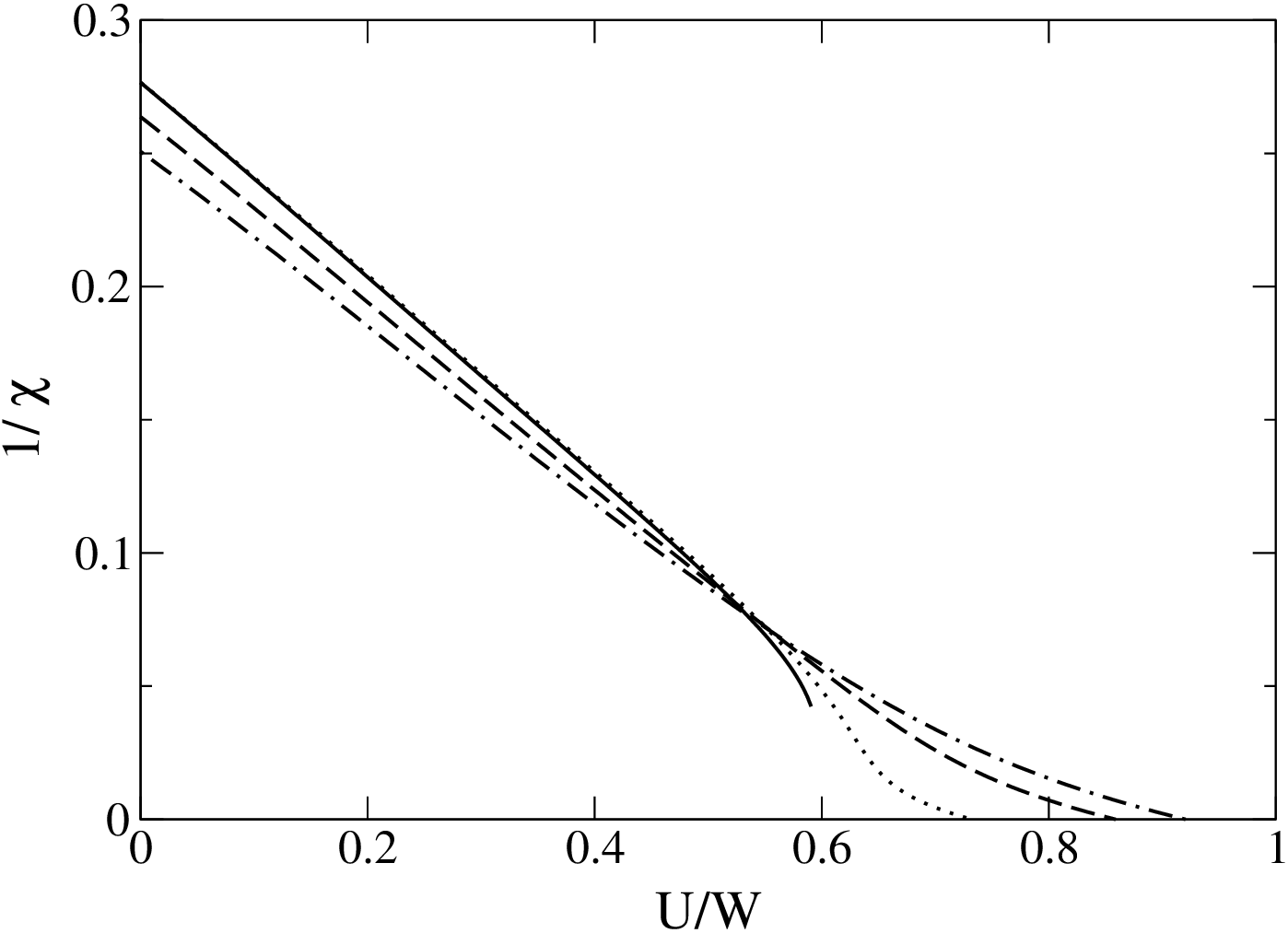}}
\caption{Inverse magnetic susceptibility for $J_H/U = 0.5$ and $\rho=2$
(full line), 2.1 (dotted line), 2.2 (dashed line), and
2.3 (dashed-dotted line).}  
\label{fig:invsuscJp5}
\end{figure}
Even though the curves are now smooth, 
one observes that the jumps occurring in the small doping regime are 
merely replaced by a very sharp drop which is, roughly speaking, 
separating a good metallic region from a nearly insulating one. 
 
We now switch to the magnetic susceptibility. The latter is obtained 
by evaluating the Gibbs free energy $G(m)$ at $m_1 = 0.1$ and $m_2 = 
0.01$, and expressing the magnetic susceptibility $\chi$ as :  
\begin{equation}\label{defchi}
\chi = \frac{m_1^2  - m_2^2}{2(G(m_1)-G(m_2))}
\end{equation} 

As shown on Fig. \ref{fig:invsuscJp5} for $J_H/U = 0.5$ the spin 
susceptibility does not diverge at half-filling, regardless of the 
value of $J_H$, though it is strongly renormalized at the Mott 
transition, where the curve stops. In the "metallic" regime, as
defined above, the 
renormalization is slightly but  increasingly suppressed with
increasing electron doping. 
%On the contrary, the renormalization is
%getting stronger with increasing  electronic concentration in the
%"insulating" regime.   
Nevertheless, in this metallic regime, there is no divergence of $\chi$,
but merely a renormalization of it by at most a factor five. 
\begin{figure}[t]
\vspace*{7mm}
\centerline{\includegraphics[height=2.5in]{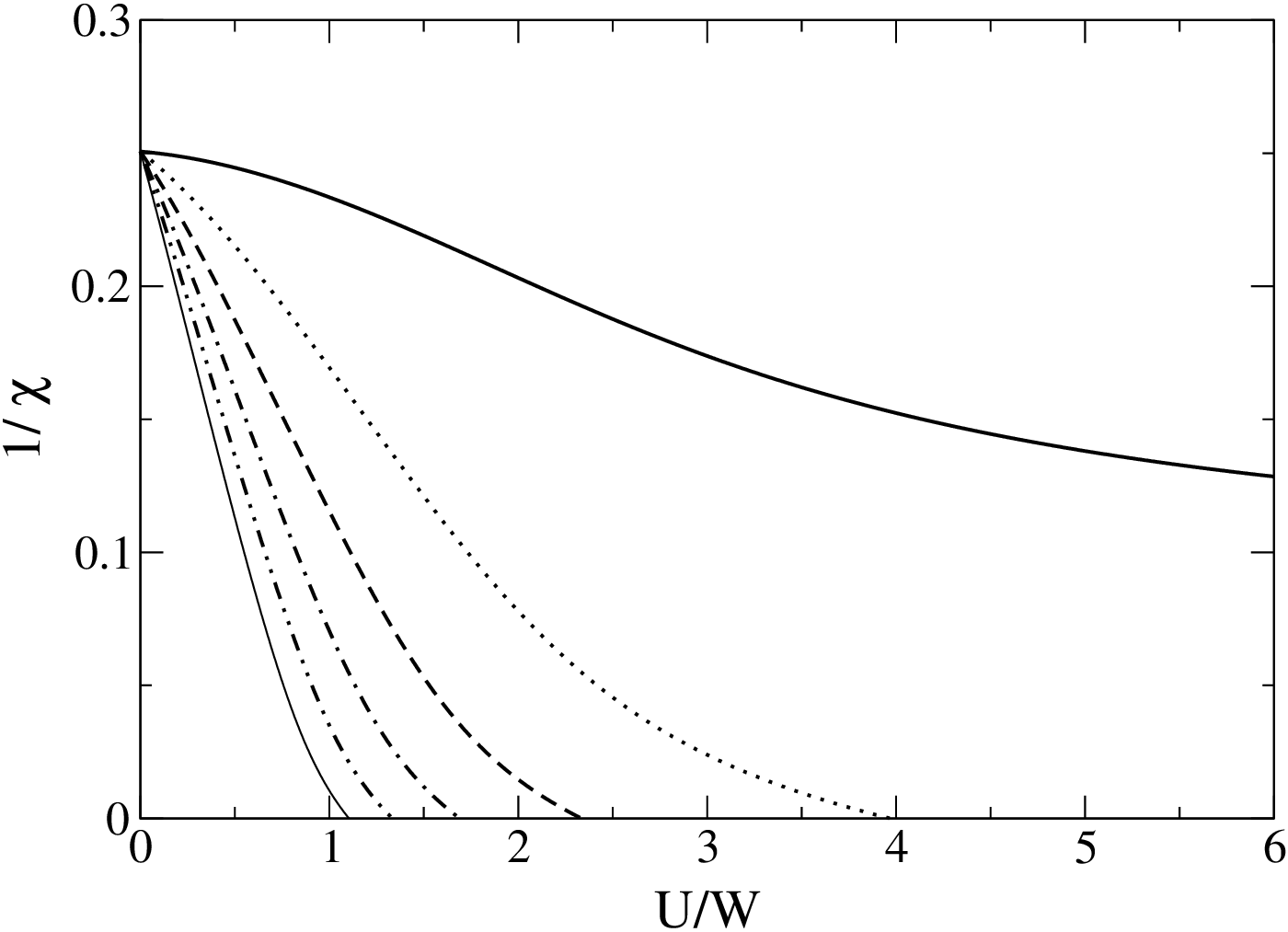}}
\caption{Inverse magnetic susceptibility for $\rho = 2.3$ and $J_H/U = $
0 (full line), 0.1 (dotted line), 0.2 (dashed line),
0.3 (dashed-dotted line), 0.4 (dashed-dotted-dotted
line), and 0.5 (thin full line). }
\label{fig:invsuscn2p3}
\end{figure}
This
renormalization is comparable to the one of the effective mass, though
somewhat larger.  

As one might anticipate, the Hund's rule coupling has a strong
influence on the magnetic susceptibility, as shown on
Fig. \ref{fig:invsuscn2p3}, where we display the inverse
magnetic susceptibility as a function of the interaction
strength for several values of $J_H/U$, at fixed density $\rho
= 2.3$. For $J_H = 0$, $\chi$ is barely affected by the
interaction, and is not found to diverge for $U$ up to 10~$W$.
Diverging susceptibilities start to appear once the
Hund's rule coupling is finite. The critical value of $U$
decreases with increasing ratio $J_H/U$, but always stays
above $U_c$. We therefore have no evidence for a weakly
renormalized ferromagnetic system. This is in agreement with the
one-band Hubbard model, where a ferromagnetic ground state
only takes place for very large interaction strengths and
moderate hole doping\cite{Moller93}. 
%This is in contrast to
%the $t-t'-U$ one-band model, where ferromagnetism may appear for weak
%coupling\cite{Muller-H95,Fresard98}.  

\begin{figure}[t]
\vspace*{7mm}
\centerline{\includegraphics[height=2.5in]{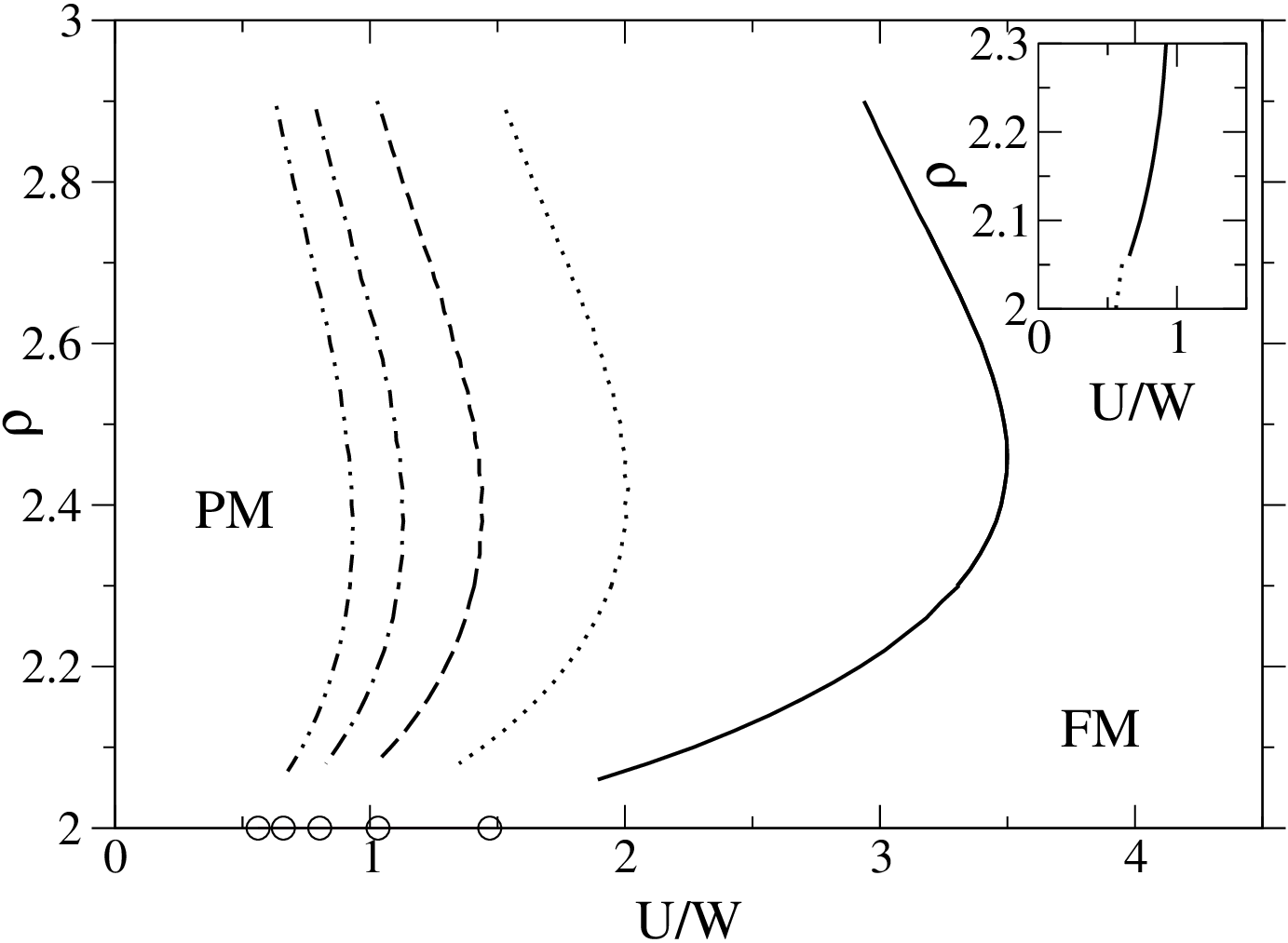}}
\caption{Instability line towards ferromagnetism in the density-U
plane for $J_H/U$=
0.1 (full line), 0.2 (dotted line), 0.3 (dashed line),
0.4 (dashed-dotted line), and 0.5 (dashed-dotted-dotted
line). The circles are indicating the corresponding $U_c$'s at
half-filling. Inset: Instability line (full line), and first order
transition line (dotted line), for $J_H/U$=0.5.}  
\label{fig:phasebou}
\end{figure}

The instabilities of the paramagnetic phases are collected on
Fig. \ref{fig:phasebou}, for several values of $J_H/U$. 
The range of stability of the paramagnetic phase is seen to depend
weakly on density for large ratio of $J_H/U$. In contrast, it may
extend to large interaction strengths for $J_H/U = 0.1$. On top there
is a strong asymmetry around $\rho = 2.5$, which is more a consequence
of the difference between $U_c$ for $\rho = 2$ and $\rho =
3$ (see Ref. \onlinecite{Lu94}), than a
consequence of neglecting four-fold occupancy. The choice
$m_1=0.1$ in Eq. (\ref{defchi}) results into a somewhat overestimated
stability range for the paramagnetic phase, but
numerical instabilities appear when working with smaller values for
$m_1$. As displayed in the inset of Fig.~\ref{fig:phasebou}, the
instability lines connect to the first order 
transition line separating two paramagnetic solutions, where the
latter ends, within numerical accuracy. No ferromagnetic solution with
magnetization $m_1$ has been found for very small doping and $U>U_c$.

When comparing this
phase diagram to La-doped Ca$_2$RuO$_4$, we see that a small amount of
electron doping turns a Mott insulator into a ferromagnet, in
agreement with experiment\cite{Cao00}. 
However the shape of the
instability lines does not appear to be compatible with the
experimental situation met in La-doped Sr$_2$RuO$_4$. Indeed no
instability line extends to the very vicinity of half-filling in the
metallic regime. This point may be easily overcome by noticing that
the band involving the d$_{xy}$ orbital is not completely
filled\cite{Singh95,Noce99}, but only carries about 1.8 electrons per
Ru atom. One should therefore concentrate on the vicinity of
$\rho=2.2$. But then, increasing the electron density from $\rho=2.2$
at fixed ratio $U/W$ does not allow for going from a paramagnetic
state to a ferromagnetic one.

In order to reconcile the theoretical and experimental data, one would
therefore need to assume that La doping not only changes the
electronic density, but also strongly influences (reduces) the band
width. An argument pointing towards this direction is provided by a
recent structural study by Friedt {\it et al.}\cite{Friedt01}, which
clearly points out that the structural parameters of the ruthenates
may easily vary to a large extend, even under a temperature
change. There is little doubt that this holds for the energy levels
$\epsilon_{d}$ and $\epsilon_{p}$  and the overlap integrals, and
therefore for the band width. Recalling that the ionic radius of
La$^{3+}$ and Ca$^{2+}$ are very similar, one may expect that La
doping indeed induces a reduction of the band width, since Ca
substitution does it, leading to a ferromagnetic instability. 
\begin{figure}[t]
\vspace*{7mm}
\centerline{\includegraphics[height=2.5in]{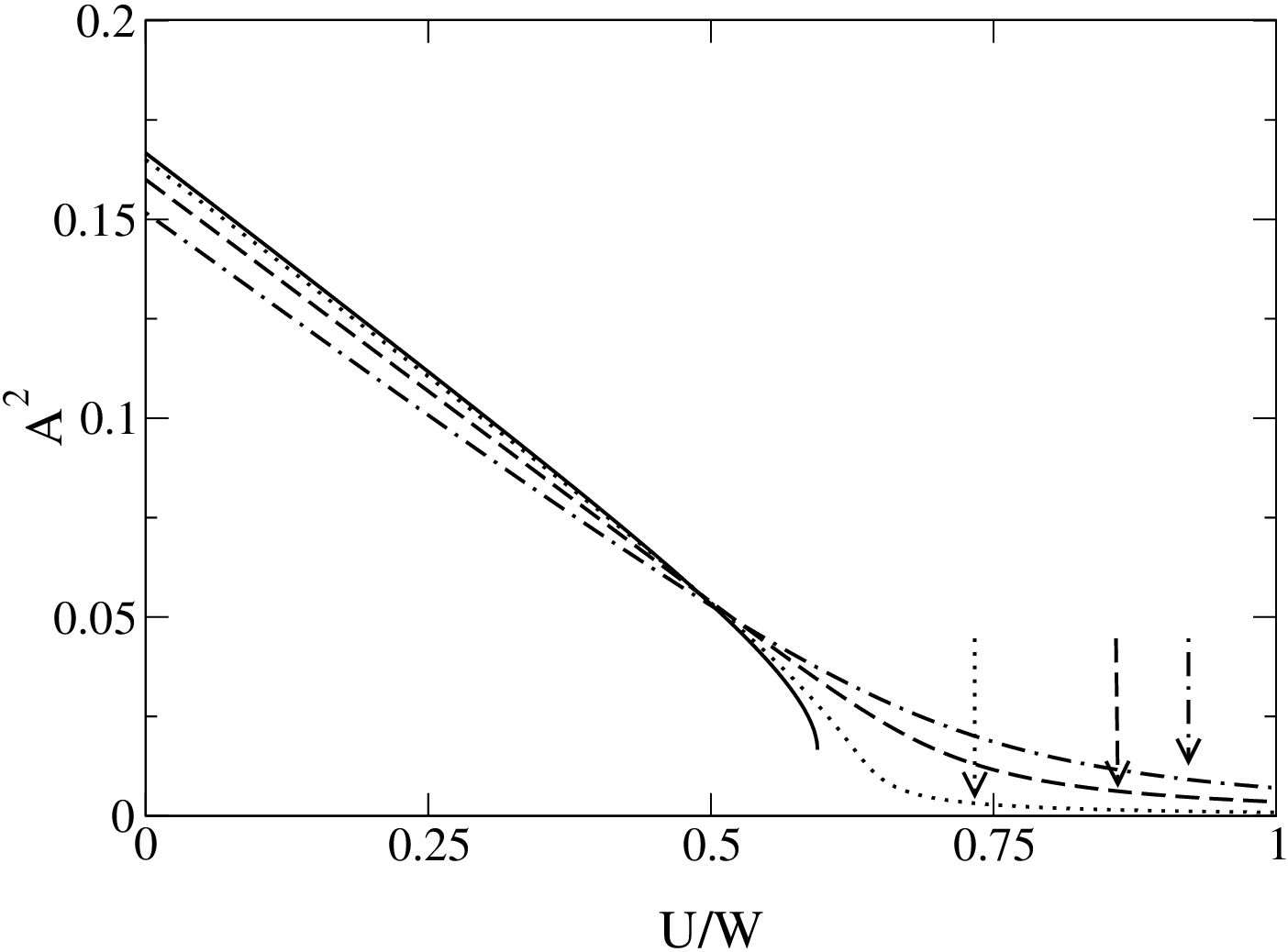}}
\caption{Double occupancy in the same orbital for $J_H/U$ = 0.5 and $\rho=2$
(full line), 2.1 (dotted line), 2.2 (dashed line), and
2.3. The arrows indicate the location of the corresponding
ferromagnetic instabilities.} 
\label{fig:boson_a}
\end{figure}
Another
reason for the discrepancy may follow from considering the d$_{xy}$
band for self-doping only, instead of including it in its own right
into the calculation from the outset. According to a recent work by
Anisimov {\it et al.}\cite{Anisimov01}, it indeed seems to play an
important role. 

\begin{figure}[t]
\vspace*{7mm}
\centerline{\includegraphics[height=2.5in]{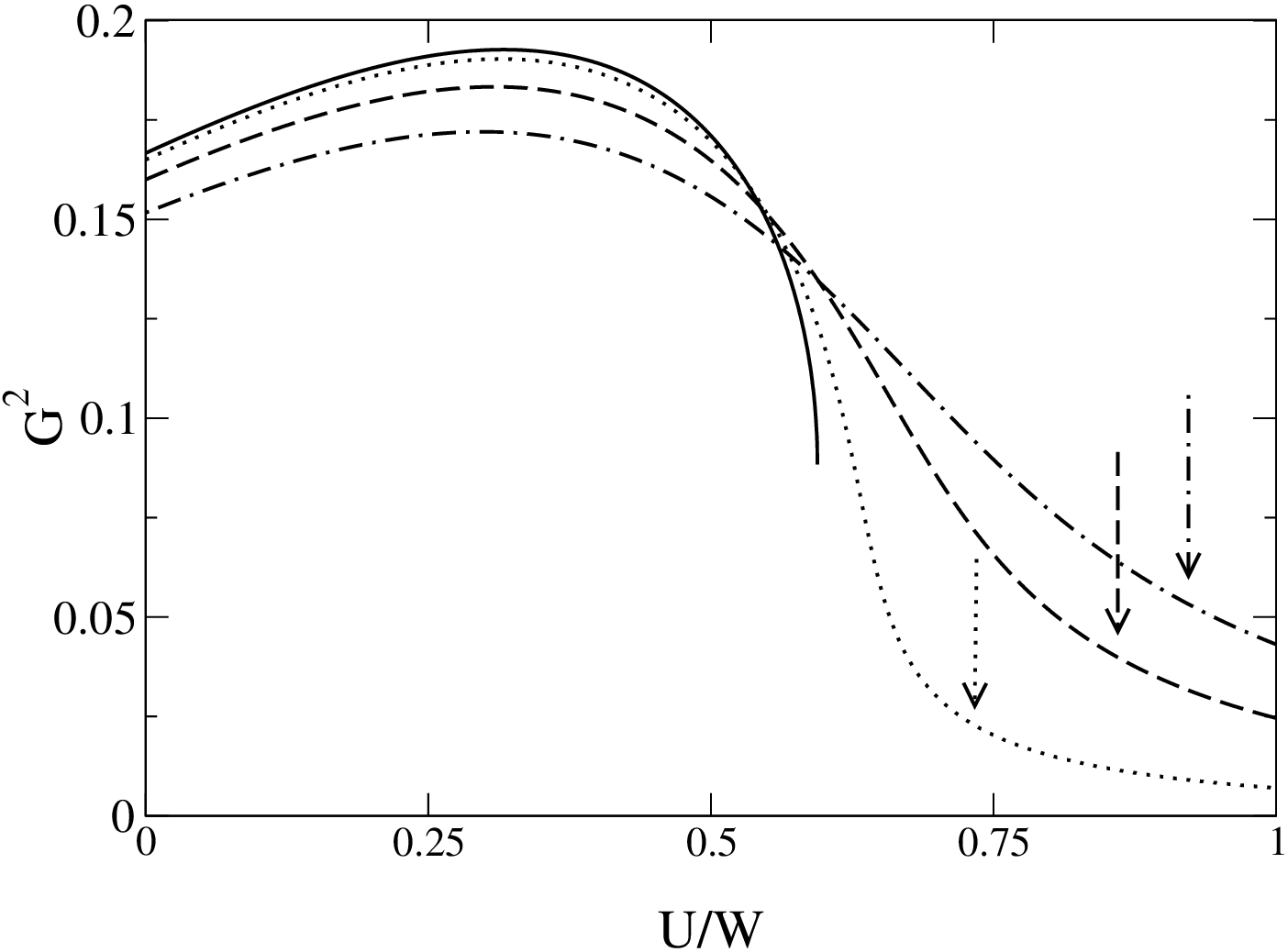}}
\caption{Double occupancy in different orbitals with opposite spin
projections for $J_H/U$ =
0.5 and $\rho=2$ (full line), 2.1 (dotted line), 2.2 (dashed line), and
2.3. The arrows indicate the location of the ferromagnetic instabilities.}  
\label{fig:boson_g}
\end{figure}

Since the four-fold occupancy turned out to be negligible, one may
wonder whether 
further simplifications could be introduced without qualitatively
altering the results. Indeed the determination of the saddle-point,
even in the simplest paramagnetic approximation is not an easy task,
and, except for the non-interacting and $J_H = 0$ limits, quite
careful numerical work is required. To address this question, we are
displaying the less probable double occupancies on
Fig. \ref{fig:boson_a} and \ref{fig:boson_g}. After introducing : $
A^2 \equiv d^2_{xz \uparrow, xz \downarrow} + d^2_{yz \uparrow, yz
\downarrow}$ and $G^2 \equiv  d^2_{xz \uparrow, yz \downarrow} +
d^2_{xz \downarrow, yz \uparrow }$, 
we see that they behave quite differently as a function of $U$ and
$J_H$. Indeed the double occupancy in the same orbital ($A^2$), which is
corresponding to the largest local interaction is suppressed
monotonically under an increase of $U$. It nevertheless keeps a
substantial value in the metallic regime, where it may therefore not
be neglected. In contrast it becomes very small in the vicinity of the
ferromagnetic instability, and presumably neglecting $A$
is not going to have a strong influence on the determination of the
instability line, especially in the large $J_H$ regime. 

The double occupancy in different orbitals with $S_z = 0$ ($G^2$) behaves
quite differently as shown on Fig. \ref{fig:boson_g}. 
An increase of
$U$ enhances it in about the entire metallic region. It then sharply
drops in the vicinity of the Mott transition, but (except for the very
vicinity of the Mott insulating state) its value remains quite
sizeable, even for strong Hund's rule interaction and large $U$. We
therefore conclude that even the instability line cannot be accurately
obtained by setting $G = 0$, in contrast to $A$. That such an
approximation, (which is corresponding to the limit $J_H \rightarrow
\infty$) will strongly influence the instability line is not
particularly amazing, since the latter does depend on $J_H$. But this
study shows that the dependence mostly arises through $G$, while the
impact of $A$ is far smaller. Therefore the precise value of the
parameter $U_3$ in Eq. (\ref{ham}) is only marginally relevant.

\section{CONCLUSION}
In summary we have investigated a two-band Hubbard model with a
realistic band structure, where no van Hove singularity appear at, or
in the vicinity of, half band filling. We have shown that the first order
metal to insulator transition that occurs at half-filling first
extends to small but finite doping in a strict sense, and second only
weakly smears out for larger doping. This divides the phase diagram
into a ``metallic'' region, where the effective mass is only weakly
renormalized and no ferromagnetic instability is found, and an
``insulating'' region, where the effective mass is strongly
renormalized, and ferromagnetism is found. On top, there is a
surprisingly small
region in-between, which is interpolating between these two behaviors.
\section*{ACKNOWLEDGMENTS}
It is our great pleasure to dedicate the present paper to Prof. Peter
W\"olfle on the occasion of his 60th birthday, especially one of
us (RF) is warmly thanking him for having been introduced to the
field of slave boson theories. We are very grateful to
V. Dobrosavljevic for having suggested this work, and for numerous
discussions, and to H. Beck for interesting discussions. We
acknowledge the financial support by the fonds national suisse de 
la recherche scientifique.

%\nocite{*}
%\bibliographystyle{ieeetr}
%\bibliography{main}

\end{document}